# ShortcutFusion: From Tensorflow to FPGA-based accelerator with a reuse-aware memory allocation for shortcut data

Duy Thanh Nguyen, Hyeonseung Je, Tuan Nghia Nguyen, Soojung Ryu, Kyujoong Lee, and Hyuk-Jae Lee, *Member*, *IEEE*

*Abstract*—Residual block is a very common component in recent state-of-the art CNNs such as EfficientNet/EfficientDet. Shortcut data accounts for nearly 40% of feature-maps access in ResNet152 [8]. Most of the previous DNN compilers/accelerators ignore the shortcut data optimization. This paper presents *ShortcutFusion*, an optimization tool for FPGA-based accelerator with a reuse-aware static memory allocation for shortcut data, to maximize on-chip data reuse given resource constraints. From TensorFlow DNN models, the proposed design generates instruction sets for a group of nodes which uses an optimized data reuse for each residual block. The accelerator design implemented on the Xilinx KCU1500 FPGA card 2.8× faster and 9.9× more power efficient than NVIDIA RTX 2080 Ti for 256×256 input size. Compared to the result from baseline, in which the weights/inputs/outputs are accessed from the off-chip memory exactly once per each layer, ShortcutFusion reduces the DRAM access by 47.8-84.8% for RetinaNet, Yolov3, ResNet152, and EfficientNet. Given a similar buffer size to ShortcutMining [8], which also "mine" the shortcut data in hardware, the proposed work reduces off-chip access for feature-maps 5.27× while accessing weight from off-chip memory exactly once.

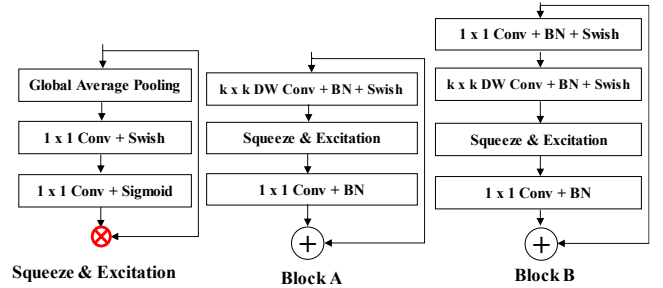

Fig. 1. Residual block with Squeeze & Excitation Optimization in EfficientNet/EfficientDet/MobileNet v3.

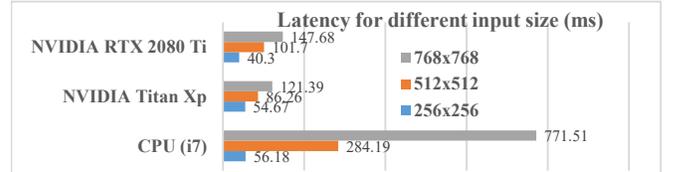

Fig. 2: Latency (ms) of the EfficientNet-B1 inference (batch size 1) for different input sizes (Tensorflow (Keras) 2.3.0, CUDA 10.2).

## I. INTRODUCTION

THERE have been many works trying to reduce the complexity and model size of CNNs using depthwise convolution [1]-[5], [40]. However, the question whether they are really efficient when running on general-purpose processor, such as CPUs/GPUs, has not been studied thoroughly. A previous study [47] showed that depthwise convolution achieves low performance on both the training and inference of various deep learning frameworks such as Tensorflow [29], Darknet [24], and Pytorch [30].

Recent state-of-the-art compact CNNs such as EfficientNet [1], EfficientDet [2] and MobileNet v3 [40], which combine mobile inverted bottleneck (MBconv) [4] and Squeeze-and-Excitation optimization (SE block) [6] as shown in Fig. 1, have achieved a new record for high accuracy in classification/detection/segmentation tasks while being less complex (i.e., BFLOPS) and more compact compared to previous works. For instance, EfficientNet-B1 achieved a higher accuracy than that of ResNet152 [22] (78.8 vs 77.8) with 7.6 times less parameters and 16 times less FLOPS. Despite being much more compact, its inference speed on an Intel CPU, NVIDIA Titan Xp (12 TFLOPS) and NVIDIA GTX 1080 Ti (11.3 TFLOPS) is not really fast on Tensorflow as shown in Fig. 2. For running on an edge accelerator such as the Google edge-TPU, the original EfficientNets are optimized by replacing depth-wise convolutions by normal convolutions and removing SE blocks at the cost of some accuracy loss [41]. Due to the very deep architecture and lightweight model, beside the kernel scheduling overhead, memory bottle-neck is also an important factor to the inference of compact CNNs. For example, for a 768x768 input size and 8-bit precision, EfficientNet-B1 requires 13.34 BFLOPS and 475 MB of intermediate data access when the inputs/outputs are accessed from the off-chip memory exactly once while the model size is merely 9 MB.

This work was partly supported by LG Display Company Ltd. and by the R&D Program of MOTIE/KEIT (No. 20010582, Development of deep learning based low power HW IP design technology for image processing of CMOS image sensors).

D. T. Nguyen, H.-S. Jae were with Department of Electrical and Computer Engineering, Seoul National University, Seoul 08826, Korea. D. T. Nguyen is now with Samsung Electronics, Korea. H.-S. Jae is now with Kakao, Korea. (e-mail: {thanhnd, hynsng}.capp.snu.ac.kr).

T. N. Nguyen, and H.-J. Lee are with the Inter-University Semiconductor Research Center, Department of Electrical and Computer Engineering, Seoul National University, Seoul 08826, Korea (e-mail: {nghiant, hyuk_jae_lee}@capp.snu.ac.kr).

S.-J Ryu is with SK Telecom, Korea (e-mail: s.ryu@sk.com).

Kyujoong Lee is with the Department of AI Convergence Sungshin Women's University, Seoul, Korea (e-mail: kyujoonglee@sungshin.ac.kr).

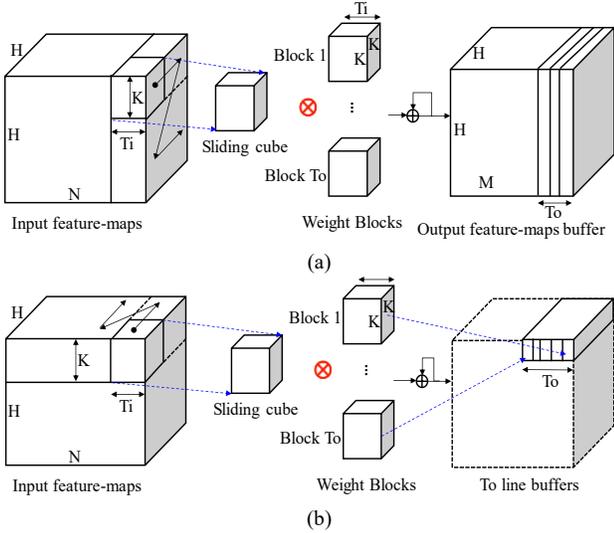

Fig. 3. The scheduling for streaming convolutional layer: (a) Frame-based weight reuse. (b) Row-based weight reuse. Borrowed from Fig. 2 in [23].
TABLE I
COMPARISON OF SCHEDULING SCHEMES FOR PIPELINE PROCESSING

| Features | Frame-based weight reuse | Row-based weight reuse |
|---|---|---|
| Input buffer size | $2 \times H^2 \times N \times Q_A$ | $(K+1) \times N \times H \times Q_A$ |
| Output buffer size | $T_o \times H^2 \times Q_S$ | $T_o \times H \times Q_S$ |
| Weight reads | 1 | H |
| Weight reuses | $H^2$ | H |
| Suitable for layer | Deep layers | Shallow layers |

N, $Q_A$: Number of input channels and bit width. H: feature-map size, $Q_S$: bit width of accumulations.

factors. An efficient scheme needs to read the weights and feature-maps exactly once from the off-chip memory with a limited on-chip buffer.

The contributions of this paper are listed below.

- An end-to-end design flow from Tensorflow frozen model to CNN inference on an FPGA-based accelerator. A CNN compiler with reuse-aware static memory allocation which supports cross-layer shortcut reuse to overcome the challenge of latency optimization with on/off-chip memory constraints for a wide range of CNNs.
- A hardware accelerator architecture with shared MAC (Multiplication-and-ACcumulation) arrays that tailors the CNN compiler is presented.
- Comprehensive experiments demonstrate that ShortcutFusion is more efficient in reducing the off-chip access with a similar buffer size compared to the previous works [8]. Even though ShortcutFusion is validated with FPGA-based accelerator in this study, ShortcutFusion is also applicable to ASIC design with a unified buffer to optimize both on-chip buffer size and off-chip DRAM access. It outperforms the recent CPUs/GPUs when running recent state-of-the-art SE-based CNNs such as EfficientNet/EfficientDet/MobileNet v3.

## II. BACKGROUND

Low latency with small batch size is very important in real time CNN inference. Therefore, this work optimizes the latency with batch size of 1. There are only two main weight block reuse schemes for the tiled-based convolutional computation: frame-based weight reuse and row-based weight reuse [23]. It should be noted that the weight reuse term here refers to each tiled weight block. For example, for a 3×3 CONV layer, each tiled weight block is $3 \times 3 \times T_i \times T_o$, where $T_i$ and $T_o$ are the parallelism factors for input and output channels, respectively.

Fig. 3(a) illustrates the computation flow of the frame-based weight reuse scheme. Because each weight block is reused for an entire frame (i.e., width × height), it needs to be read from the buffer exactly once. Therefore, only a small weight block needs to be buffered. It is noteworthy that the input/output feature-maps are accessed multiple times from the on-chip buffer. This scheme is efficient for layers with a large weight size and a small feature-map size, in which the input/output feature-maps can be accessed on-chip. Meanwhile, the latency of reading the weight blocks from the off-chip memory can be hidden by the computation of the sub-frame input.

The computation of the row-based weight reuse scheme is depicted in Fig. 3(b). It should be noted that the inputs are

There are previous works on end-to-end frameworks for accelerator designs on FPGAs [15]-[19], [33], [42]-[46]. FP-DNN in [15] allocates a minimal number of physical buffers in DRAM (not SRAM) as a memory pool for implementation; it does not leverage the on-chip buffer in the FPGA efficiently to reduce the off-chip access. Argus in [16] provides an end-to-end framework for a multi-CLP type accelerator [53]. The studies in [18] and [19] also provide an end-to-end framework for a multi-layer processor. Because the BRAMs of a FPGA are not enough for multiple hardware units, the data have to be stored in the off-chip memory. Therefore, in [16], [18], and [19], a large amount of data access is required for deeper networks. Even though they work fine for shallow networks, their scalability to a wide range of CNNs is limited. There are existing frameworks from Tensorflow to FPGAs, such as in [17], [33], [42] and [44]-[46]. DNNWeaver [42] does not support layer fusion while the others only support the fusion of Convolution, Activation and Batch Norm and/or Pooling. As mentioned in [8], the shortcut connection accounts for 40% of the feature-maps access in Resnet-152. Their study shows that smaller on-chip buffer size and unnecessary data access elimination translate into 24.9% reduction in FPGA power consumption for ResNet152. All of these frameworks do not support in-hardware flexible data reuse schemes and neglect cross-layer shortcut data optimization which might cause sub-optimal off-chip access.

There have been many previous studies on the dataflow of CNN computing [8]-[12], [51]. Among [9], FlexFlow [11] [12], and [51] which optimize the dataflow for a shallow network VGG16, SmartShuttle [12] achieves the highest number of MAC/DRAM accesses. It proposes to switch between two data reuse schemes: partial sum oriented and weight oriented. It requires 434.8 MAC/DRAM accesses (i.e., 214 MB) for running the CONV layers of VGG16. For a deeper network such as ResNet-152 or EfficientNet, larger off-chip accesses might cause a long latency because it requires larger data for shortcut connection and feature-maps. As reported in the paper, the buffer size, which is larger than 512 KB, does not help to reduce the DRAM access despite supporting flexible tiling

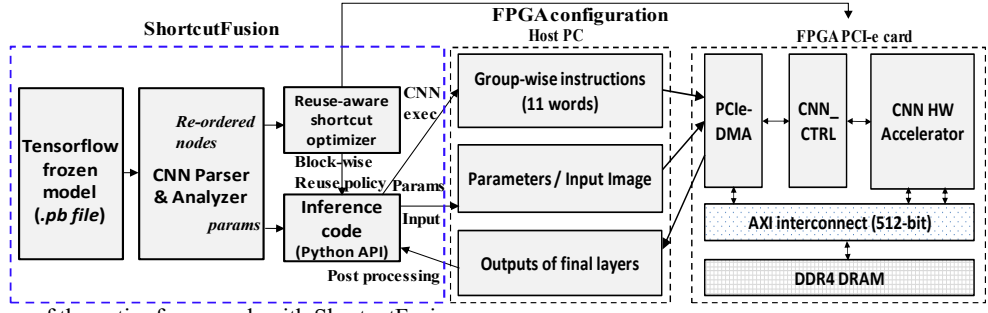

Fig. 4. Block diagram of the entire framework with ShortcutFusion.

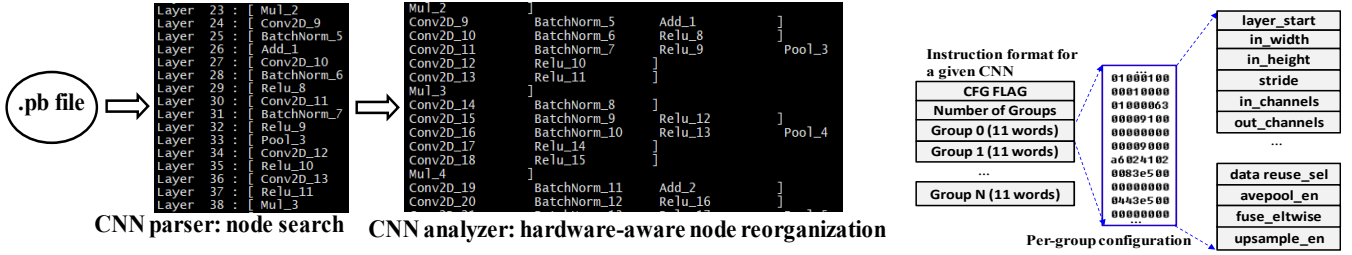

(a) EfficientNet: CNN analyzer re-organizes 418 nodes to 139 groups for execution (b) Group-wise instructions

Fig. 5. (a) Example of the CNN parser & analyzer for the EfficientNet protobuf file. (b) Instruction generation for off-loading the CNN to the FPGA.

loaded once in a row-by-row manner. The input sliding cube, while slides along the width of the input image, is convolved with $T_o$ weight blocks each time to produce $T_o$ temporary output values. These weight blocks are reused for a row. The input sliding cube then shifts $T_i$ channels toward the end of N-input channels until all input channels are read. To finish the processing for one row, the entire weight of the model is accessed from the memory. Therefore, to process the whole input feature-maps, the weights are read H times.

Table I borrowed from [23] summarizes the characteristics of the two weight reuse schemes. It should be noted that $Q_S$ is much larger than $Q_A$. Therefore, the frame-based weight reuse scheme is suitable for deep layers where the feature-maps size is small, while the row-based weight reuse scheme is very efficient at reducing buffer size for shallow layers where the feature-map size is large, and weight size is small.

There are previous works in [8], [10], and [12], which minimize off-chip data access by using tiling along width/height of inputs and mixing different data reuses: Weight Reuse (WR), Input Reuse (IR), Output Reuse (OR). The study in [12] shows that OR has more benefit than IR so they mixed OR and WR. The frame-based weight reuse scheme in Fig. 3(a) is actually the Weight Reuse scheme in [12] except that the proposed scheme reuses weights for entire frame not a tile. The proposed scheme switches between the row-based reuse (Fig. 3(b)) and frame-based reuse (Fig. 3(a)) to efficiently reduce off-chip access for shallow layers while completely remove feature-map off-chip access for deep layers.

### III. SHORTCUTFUSION

#### A. Overview

A block diagram of the proposed framework is shown in Fig. 4. There have been many open-source deep learning frameworks for deep learning research: Tensorflow [29], Pytorch [30], Caffe [31] and Darknet [24]. Tensorflow is one of the most popular DNN frameworks that was developed by Google. Hence, this paper uses Tensorflow in the front-end of ShortcutFusion.

It is well known that a CNN is tolerant to errors. Previous research in [26] and [27] shows that 8-bit is efficient for various DNN inference tasks. The Google TPU [28] also uses 8-bit for both training and inference. Regarding the EfficientNet quantization, there are previous works about 8-bit/4-bit quantization with comparable accuracy to floating point model [60], [61]. Therefore, this study adapted an 8-bit quantization for the accelerator design. The Tensorflow model file (protobuf file) is exported to the *CNN parser & analyzer* for parsing the architecture of the given CNN and extracting the quantized parameters. As depicted in Fig. 5 (a), the searched nodes of the CNN architecture are then reorganized into *groups* as supported by the back-end accelerator. Existing frameworks such as CloudDNN [17], [33], TensorRT [39] and Xilinx ML-Suite [44] support the fusion of CONV, Relu and BatchNorm and/or Pooling only. The lack of reuse of cross-layer shortcut data might cause large off-chip accesses thereby affecting the system performance. To address this issue, ShortcutFusion try to group as many nodes as possible to reduce the intermediate data movement and runtime overhead. For example, Convolution, Activation, Normalization, Pooling, Element-wise (shortcut pass), and/or Up-sampling layers are fused together as supported in the back-end accelerator. Like TensorRT, the feature-merging for concatenation in the row-reuse mode of this work also supports redirecting the output to the eventual destination of the concatenation to avoid data movement. The parameters extracted from the *CNN parser & analyzer* are used in the unified software reference code for hardware verification. In addition, the CNN architecture information is used in the *Block-wise optimizer* to select the optimal data access scheduling for each group of nodes in terms of the latency, on-chip buffer requirement and DRAM access. The on-chip buffer selection and parallelisms are taken to configure the accelerator. It is noteworthy that the *reuse-aware shortcut optimizer* satisfies a strict constraint of the DRAM access, in which the parameters are accessed exactly once, the inputs/outputs of some layers are accessed from the DRAM exactly once, and the inputs/outputs of other layers are stored on-chip. This constraint is used in optimization problem in section IV-B (equation 10). Finally, the *inference code* generates instructions for entire layers of the CNN. As depicted in Fig. 5(b), the instruction sets of each *node group* consists of 11 words describing convolution size, activation type, pooling/upsampling option, fused element-wise, etc. It is noteworthy that the *inference code* packs parameters, input and all instructions and sends them at once to the hardware accelerator. The detail of the *reuse-ware shortcut optimizer* will be presented in section IV.

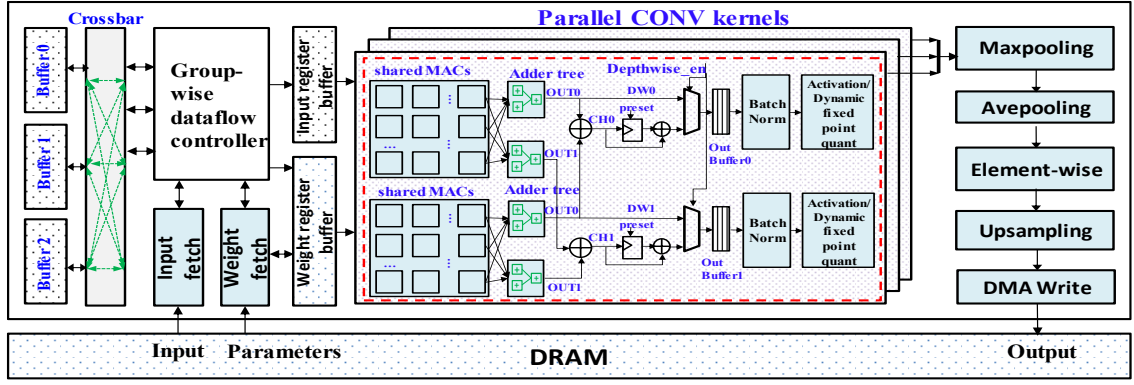

Fig. 6. Block diagram of the CNN accelerator.

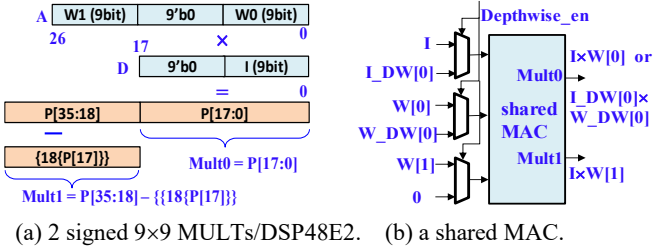

(a) 2 signed 9×9 MULTs/DSP48E2.   (b) a shared MAC.

Fig. 7. (a). Mapping of 2 signed 9x9 MULTs to a DSP48E2. (b). Shared MAC for a normal/depth-wise convolution.

### B. Architecture of the FPGA-based CNN accelerator to support ShortcutFusion

Fig. 6 describes the architecture of the accelerator that tailors ShortcutFusion. As mentioned in [8], the shortcut connection accounts for 40% of the feature-maps access in Resnet-152. Therefore, to maximize the shortcut data reuse, the proposed accelerator has an additional buffer for shortcut data. These physical buffers {0,1,2} are three interchangeable buffers for storing the input, output, shortcut data, or parameters of the entire layer. The two reuse schemes are valid for both normal and depthwise convolutions. Therefore, the group-wise dataflow controller is able to switch between two levels of weight reuse based on the per-group instructions to balance between computation and off-chip memory access. The wide circular *row buffer* and double weight buffer provide the high bandwidth for feeding the sliding windows and weight to the parallel convolution kernels. The partial sum from the MAC arrays are stored temporarily in the *out buffer*. As different CNN layers may require different value ranges for data, the proposed design supports a dynamic fixed point format to preserve the accuracy. It should be noted that swish and sigmoid activation only support a single fixed point format due to the hardware overhead. Swish and sigmoid activations are implemented using an 8-bit look-up table. Two look-up tables share a single 18Kb Block RAM. Therefore, a total of $T_o$ 18Kb Block RAMs are required. Finally, there is a chain of modules such as max-pooling, average-pooling, element-wise addition, and up-sampling. The outputs from the parallel kernels are forwarded directly to this chain without storing back to the memory to reduce the data movement. It should be noted that these modules also support different data reuse schemes same as the convolutional module, thereby, have connection to the three physical buffers.

*1) Convolution kernel design with shared MACs*

A single DSP48E2 in Ultrascale and Ultrascale+ supports two INT8 multiplications sharing a same operand [32] to increase the DSP efficiency. As the proposed CNN accelerator targets multiple CNNs in various applications, it supports both 8-bit signed and unsigned feature-maps. In addition, the weights use 8-bit non-zero quantization which has been proved to have a higher accuracy compared to the zero-centered quantization [23]. Therefore, the proposed design requires signed 9x9 multiplication which is not inherently supported by DSP48E2 in the double MAC mode. To make it possible, a correction logic is added as described in Fig. 7(a), where $Mult0 = I \times W_0$, $Mult1 = I \times W_1$.

Double multiplication can be applied to the normal convolution because the input feature-maps are shared among the different weight filters. However, the depth-wise convolution does not have such input reuse. Fig. 7(b) shows the block design of a shared MAC. It supports double multiplications for the normal convolution and single multiplication for the depth-wise convolution. Finally, as depicted in Fig. 8, this study proposes a convolution kernel design with a *shared MAC array* to utilize the DSPs better. Fig.

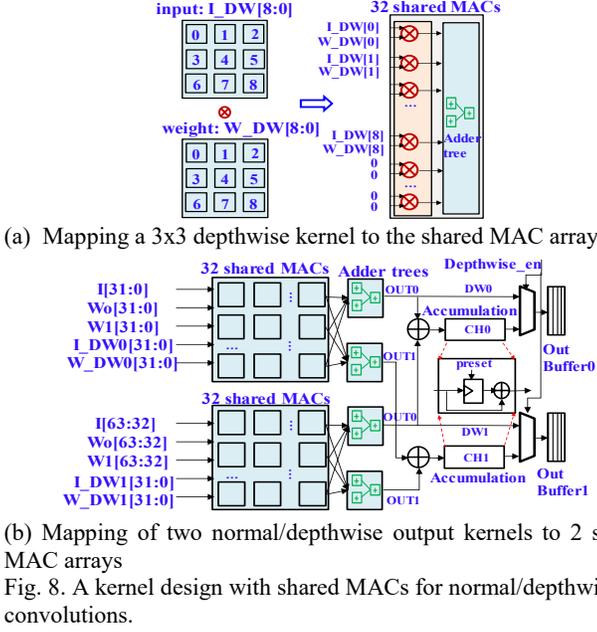

(a) Mapping a 3x3 depthwise kernel to the shared MAC array

(b) Mapping of two normal/depthwise output kernels to 2 shared MAC arrays

Fig. 8. A kernel design with shared MACs for normal/depthwise convolutions.

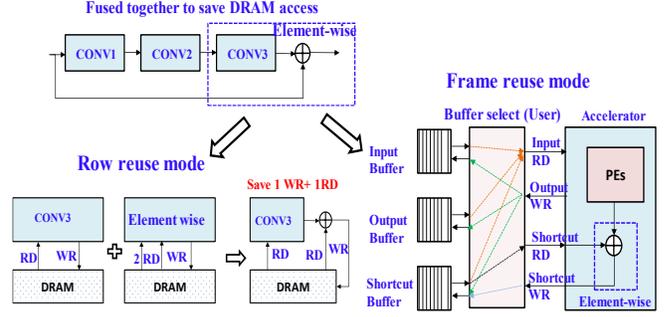

Fig. 9. Shortcut data reuse in row-reuse mode (left) and frame-reuse (right).

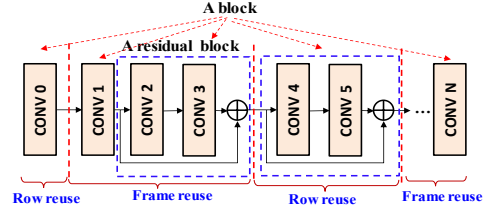

Fig. 10. Block-wise data reuse switching.

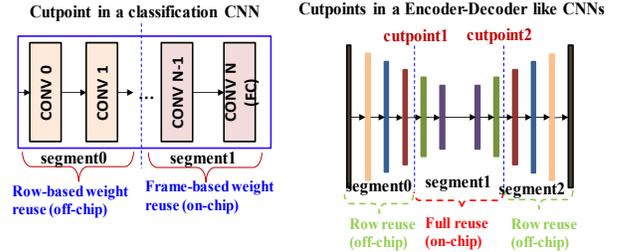

Fig. 11. Examples of a single cut-point (left) and double cut-points in CNNs.

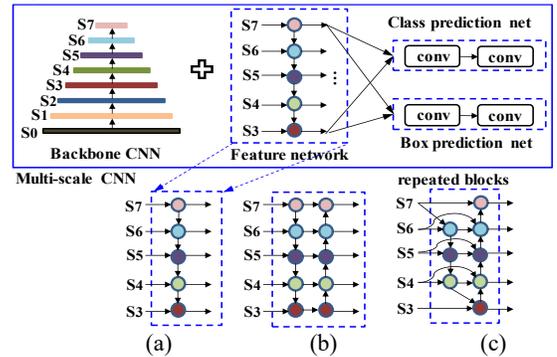

Fig. 12. Categorization of CNNs according to the architecture of the Feature Network. (a) FPN [34]. (b). PANet [35]. (c). BiFPN [2].

8(a) depicts the mapping of a 3×3 depth-wise kernel to the MAC array. Because the recent popular CNNs such as MobileNet v3, EfficientNet and EfficientDet use 1×1/3×3/5×5 depth-wise kernels, the MAC array is able to process a kernel in one cycle. In the case where the kernel size is greater than 7×7, it needs multiple cycles for kernel processing. Fig. 8(b) shows the detailed mapping of two output kernels computation to the two MAC arrays. Sixty-four multiplications for each normal CONV kernel are interleaved into two *shared MAC* arrays, $I[63:0] \times W_{\{0;1\}}[63:0]$, whereas each depth-wise kernel is processed by separated MAC arrays: $I\_DW_0[31:0] \times W\_DW_0[31:0]$ (top array), $I\_DW_1[31:0] \times W\_DW_1[31:0]$ (bottom array). Depending on whether it is the normal or depth-wise CONV, the output from the adder tree is forward to the accumulation unit or output buffer directly. It is noted worthy that the shared MAC array contains 2048 MACs, which supports 4096 multiplications per second for normal convolution and 2048 multiplications per second for depthwise convolution. Regarding the adder trees, each convolution kernel contains four 32-input binary adder trees. The total adder tree numbers are thus 256.

*2) Shortcut data reuse in different weight reuse schemes*

In order to realize ShortcutFusion, the back-end hardware need to support cross-layer shortcut data reuse. As illustrated in Fig. 9, the second inputs of the element-wise addition layer (shortcut layer) are fetched whenever the first inputs from the convolution kernels are available. Therefore, in the row-reuse mode, the fused layers (CONV+shortcut) require only one Write and two Reads instead of two Writes and three Reads from the off-chip memory. Similar to the row-reuse mode, the frame-reuse mode uses two less on-chip data movements. Moreover, the element-wise layer does not incur an additional timing overhead, thereby reducing the total latency.

ShortcutMining [8] reuses shortcut data on-chip by reserving an untouched buffer space for shortcut data. Since it uses a fixed reuse scheme for all layers, it required a large buffer size. On the other hand, the proposed scheme carefully selects weight reuse scheme in the memory allocation for shortcut layer data, thereby very is efficient in reducing both the total latency, off-chip access, and on-chip buffer size.

IV. REUSE-AWARE SHORTCUT OPTIMIZER

Because this study supports shortcut data reuse for different reuse schemes, the proposed scheme is able to switch the data reuse between blocks of CONV layers called block-wise data reuse. As shown in Fig. 10, a block of layers is defined as a residual block or a single CNN layer which does not belong to any residual blocks. Given the buffer constraints, the *proposed optimizer* searches for the optimal switching between two weight reuse schemes (row-reuse and frame-reuse) for each block to get the optimized latency, on-chip buffer, and DRAM access. It should be noted that both the weights and feature-maps are read from the off-chip memory exactly once.

A cut-point is defined as the position in the CNN graph where the data reuse scheme switches. A CNN comprised of N basic blocks might have up to $2^N$ different switching schemes. The exhaustive search to find the optimum data reuse policy is impractical for general CNNs which have hundreds of layers. Moreover, given a buffer constraint, it is not possible to get an optimized reuse policy for all blocks because each block requires a different constraint for its distributed buffers. There is an observation that, in all the recent CNNs, the feature-map size monotonically increases or decrease in a certain sequence of blocks. Therefore, this study proposes a coarse-grained block-wise shortcut reuse scheme which has been validated by recent very deep CNNs. In the proposed relaxation, a sequence of increasing or decreasing size blocks is assumed to have exactly one cut-point. Cut-points divide a CNN into segments as illustrated in Fig. 11. Layers in a same segment uses the same weight reuse scheme. The number of cut-points depends on the CNN architecture varying from a plain structure or residual style to a multi-scale, multi-branch architecture. In Fig. 11, a classification CNN has a single cut-point because the CNN structure goes from the largest scale to the smallest scale. With the same intuition, an auto-encoder CNN has two cut-points. Fig. 12 shows the details of the CNN categorization according to the Feature Network which might require a different number of cut-points. Fig. 12(a) shows the object detector with the Feature Pyramid Network (FPN [34]). Yolov3 [21] and RetinaNet [36] also use an FPN network. These CNNs require two cut-points. PANet [35] fuses the feature-maps both top-down and bottom-up. Therefore, PANet requires three cut-points. For the recent state-of-the-art object detector/segmentation EfficientDet [2], the number of cut-points depends on the number of BiFPN (Bidirectional Feature Pyramid Network) layers. For example, if the repeated block is one, there are only three cut-points because there are only one top-down and one bottom-up path aggregation. On the other hand, if there are more than one repeated block, the number of cut-points are equal to ($2 \times repeated\_blocks + 1$).

The challenge is to find the cut-point positions to achieve the minimum latency while satisfying the buffer constraints and DRAM access constraints.

### A. Reuse-aware static memory allocation

For the row-reuse mode, memory space for inputs, outputs and shortcuts are allocated in the off-chip memory. On the other hand, in the frame-reuse mode, the inputs, outputs and shortcuts are allocated to one of the three physical buffers to eliminate the off-chip access. Given the CNN architecture, the memory allocator statically allocates buffers for each layer by assigning three variables {*alloc_input, alloc_output, alloc_shortcut*} to {*0, 1, 2*} properly to reuse the shortcut data that is stored on-chip. It should be noted that the data of the long-path shortcut

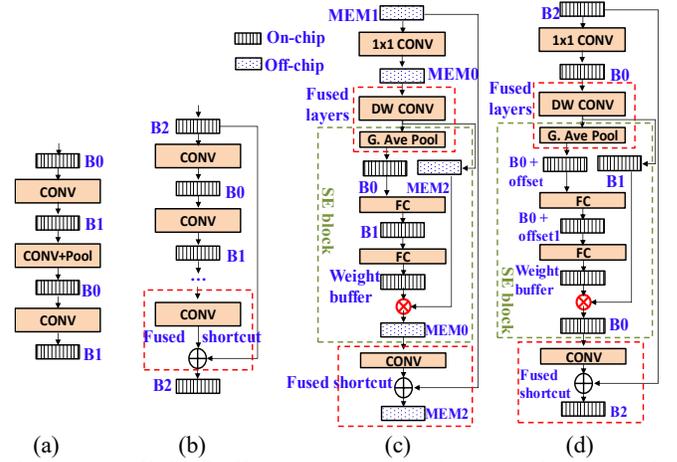

Fig. 13. On/Off-chip buffer management in ShortcutFusion. (a). Plain network. (b). Network with the residual block. (c) Residual block w/ the Squeeze-and-Excitation (SE) block in row-based weight reuse. (d). Residual block w/ the SE block in frame-based weight reuse.

connection for concatenation is stored off-chip to avoid long lifetime data in the on-chip buffers.

Fig. 13 shows detailed examples of the on/off-chip memory access management for different network structures. In Fig. 13(a), networks with a plain structure such as VGGNet, Darknet19 and SimYolov2 [20] require only two buffers. On the other hand, the CNN with the residual block in Fig. 13(b) requires three buffers to reuse the shortcut data. The outputs from the last CONV layer in a residual block are forwarded to the Shortcut layer to reduce intermediate data access.

Fig. 13(c) and 13(d) show the memory allocation for the residual block with the Squeeze-and-Excitation optimization with different weight reuse schemes. In the row-reuse mode as shown in Fig. 13(c), the outputs from Global Average Pooling and two FC layers are stored on-chip because their size is small. The last layer in the SE block is a scale layer (i.e., the red multiplier in the figure) that works in the same way as the 1x1 depthwise CONV layer without batch normalization. Different from the row-based reuse mode, the frame-reuse mode in Fig. 13(d) completely allocates data to three on-chip buffers. The outputs from the depthwise CONV layer (DW CONV) are stored in buffer B1. In parallel, the outputs from (DW CONV + Pooling) are stored in buffer B0 with an offset address to avoid overwriting the input in buffer B0. Similar to Fig. 13(c), the outputs from the two FC layers and data in buffer B1 serve as weights and inputs for the scale layer, respectively. The residual block with the SE optimization is known to be inefficient in GPUs even though it does not incur much computation overhead. This dataflow-aware static memory allocation and on-chip data forwarding are very efficient in reducing off-chip memory access for residual block with SE optimization.

### B. Optimizing shortcut data reuse with given constraints

Let us denote that *L* is the data reuse policy. It is noteworthy that layers in a same basic block use the same data reuse. To calculate the required size for each buffer {0,1,2} according to the data reuse *L*, the static buffer allocation for each layer *i* needs to be considered as shown at step 1 in Algorithm 1.

**Algorithm 1: Calculation of the required buffer size.**

```
Input: CNN architecture with N layers
Output: required buffer size w.r.t. the data reuse
policy L: buff[0](L), buff[1](L), buff[2](L)

For each layer i in frame-reuse do
 1. {alloc_in(i), alloc_out(i), alloc_shortcut(i)}
      = buff_alloc(layer i);
 2. buff[alloc_in(i)](L) =
      max(buff[alloc_in(i)](L), input_size(i));
 3. if (to_residual(i) == yes) // layer i is used for
    residual layer
      buff[alloc_shortcut(i)](L) =
       max(buff[alloc_shortcut(i)](L),output_size(i));
 4. if (NextLayer(i) == Maxpool2x2) // fused conv+pool
      buff[alloc_out(i)](L) =
       max(buff[alloc_out(i)](L), output_size(i)/4);
    else
      buff[alloc_out(i)](L) =
       max(buff[alloc_out(i)](L), output_size(i));
End_for
```

{*alloc_in(i), alloc_out(i), alloc_shortcut(i)*} are the buffer allocations for the input/output/shortcut data of layer *i*, respectively.

To derive the total raw SRAM size, the size of the following buffers need to be calculated: *weight_buff, row_buff, out_buff,* and *write_buff*. It is noteworthy that all the buffers have the same number of banks which are the parallelism factors $T_i=T_o$ to remove the logic congestion of the buffer bank selection. In the row-reuse mode, the entire weights of a layer are pre-loaded to the on-chip buffer. Therefore, the required buffer size for a weight is as follows:

$$weight\_buff(L) = max_{i=row\_reuse} weight\_size(i) \quad (1)$$

It should be noted that the double weight block buffer for feeding weight blocks to parallel convolution kernels is stored in the LUT-RAMs of the FPGA chip because its depth is small (2×3×3 = 18). Because buffer 1 is shared for both feature-maps and weights, the size of buffer 1 is as follows:

$$buff[1](L) = max(weight\_buff(L), buff[1](L)) \quad (2)$$

The proposed convolutional kernel design focuses on the 3x3 normal convolution as in most of the CNNs and 1x1/3x3/5x5 depthwise convolutions for EfficientNet/Det. However, it can also support convolutions with a filter size larger than 7x7 by merely increasing the number of row buffers and double weight block buffer depth. Therefore, in the proposed design, there are six rows in the row buffer including one row for input prefetching:

$$row\_buff(L) = max_i 6 \times in\_rowsize(i) = max_i 6 \times w_i \times N_i \quad (3)$$

where $w_i$ and $N_i$ are the width and input channels, respectively. Regarding the buffer for temporarily partial sums, the buffer size for the frame-based reuse mode is larger than that for the row-based reuse mode because the row-reuse mode needs to buffer only one row. Therefore, the partial sum buffer (4-byte width) is derived as below:

$$out\_buff(L) = max_{i=frame\_reuse} out_{buff(i)}$$
$$= max_{i=frame\_reuse} out\_w_i \times out\_h_i \times T_o \times 4 \quad (4)$$

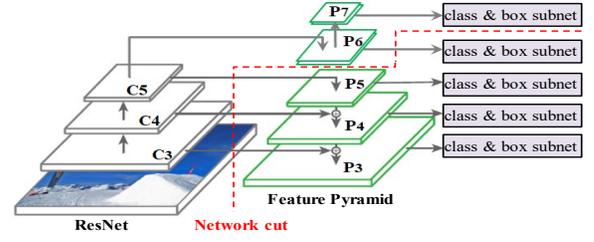

Fig. 14. RetinaNet with a network "cut."

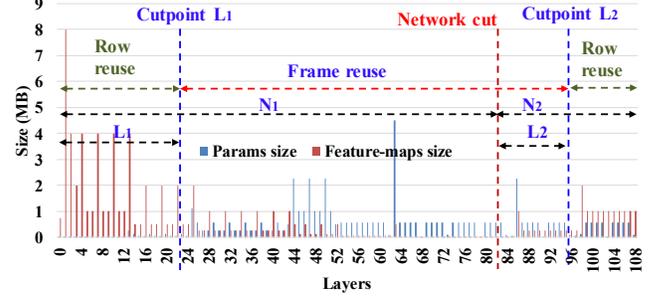

Fig. 15. Double cut-points (L$_1$, L$_2$) in RetinaNet.

Finally, the output from the accelerator is buffered in the write buffer before writing to the off-chip memory. In the frame-based reuse mode, except for the final outputs of the CNN and long-path shortcut/concatenation, all intermediate data are stored on-chip. Therefore, the size of the write buffer is as below:

$$write\_buff(L) = max(max_{i=row\_reuse} write\_buff(i),$$
$$max_{i=frame\_reuse \& i=final\_layers} write\_buff(i))$$
$$= max(max_{i=row\_reuse} out\_w_i \times T_o,$$
$$max_{i=frame\_reuse \& i=final\_layers} out\_w_i \times out\_h_i \times T_o) \quad (5)$$

To sum them up, the required raw SRAM size is as follow:

$$SRAMsize(L) = row\_buff(L) + out\_buff(L)$$
$$+write\_buff(L) + buff[0](L) + buff[1](L) + buff[2](L) \quad (6)$$

The raw SRAM size does not physically reflect the real BRAM utilization of a FPGA chip. Therefore, the number of BRAM 18k needed for each buffer is estimated as below:

$$BRAM18k = \left\lceil \frac{buff\_depth}{1024} \right\rceil \left\lceil \frac{buff\_width}{18} \right\rceil \quad (7)$$

Regarding the necessity of reducing the total DRAM access in a CNN computation, the previous study in [37] shows that the energy consumed by an off-chip access is much larger than that by an on-chip access or arithmetic operation in an ASIC chip. Therefore, it is important to constrain the off-chip access. The proposed design supports data forwarding as discussed in section III-B-2. Hence, the DRAM access for feature-maps, and total DRAM access are calculated as shown in (8), (9), respectively.

$$DRAM\_FM(L) = \sum_{i=row\_reuse, i=conv} (in\_size(i) + out\_size(i))$$
$$+ \sum_{i=row\_reuse, i=shortcut} in\_size(i)$$
$$+ \sum_{i=frame\_reuse, i=concat/route} 2 \times in\_size(i) \quad (8)$$

$$TotalDRAM(L) = DRAM\_FM(L) + \sum_i weight\_size(i) \quad (9)$$

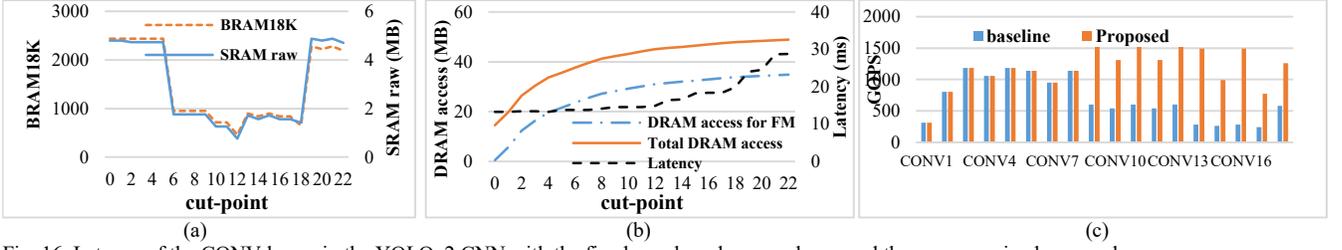

Fig. 16. Latency of the CONV layers in the YOLOv2 CNN with the fixed row-based reuse scheme and the coarse-grained reuse scheme.

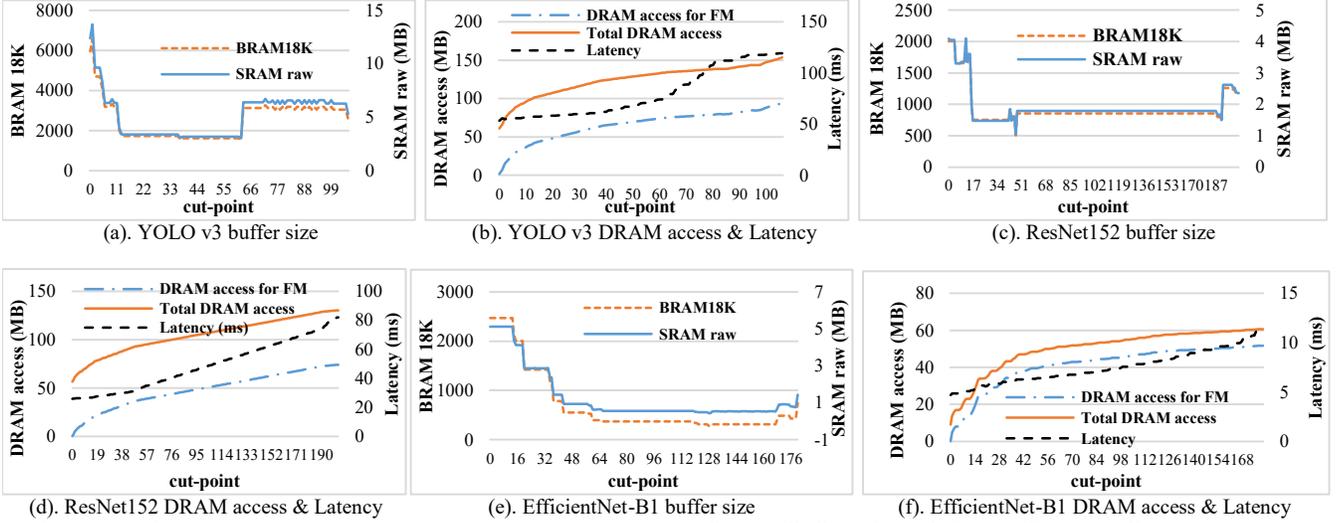

Fig. 17. The on/off chip access and latency w.r.t. the cut-point position for various CNNs: YOLO v3, ResNet152, and EfficientNet-B1.

The optimization problem is to find the data reuse policy $L$ as follows:

$$min_l(latency(L, T_i, T_o)) \text{ s.t.:}$$
$$DSP(T_i, T_o) < \alpha \quad (*)$$
$$BRAM18K(L, T_i, T_o) < \beta$$
$$weights\ access = 1, feature\ maps\ access \leq 1 \quad (10)$$

where the constraint (10) is explained in section III-A. If different FPGA is selected, we can decide the parallelisms (i.e., MAC array size) of the accelerator and the switching points of the reuse schemes based on the optimization (*).

A problem is raised such that the latency estimation by running the RTL simulation for each candidate takes a very long time. Because the number of candidates in the design space can be large, the RTL simulation approach is not feasible. Therefore, this work built a cycle-accurate timing simulator to estimate the latency of a CNN layer running different reuse schemes as described in Fig. 3(a) and Fig. 3(b). The latency estimation model was verified with the RTL simulation for the CNN with all the candidates for policy $L$ in the design space.

*Single cut-point optimization:* In Fig. 11 (left), the *i-th* layer is in the row-reuse mode if $i < L$ and the frame-reuse mode, otherwise.

*Multiple cut-points optimization:* Fig. 14 shows an example of double cut-points in the RetinaNet. Network "cut" is the position that divides the RetinaNet into two sub-networks: from the beginning to the smallest scale and from the smallest scale to the end. From this network cut, the relative position of the data reuse policy $L = (L_1, L_2)$ is defined, as shown in Fig. 15, where $0 \leq L_1 < N_1$, and $0 \leq L_2 < N_2$. The real layer indexes of L are $(L_1, N_1 + L_2)$. Layer i=row_reuse if $i < L_1 \| i \geq N_1 + L_2$, and i=frame_reuse, otherwise. The other multiple cut-point cases can be extended with a same exhaustive search for the optimum policy in the polynomial of time $O(N^k)$, where N and k are the depth of the sub-networks and the number of the cut-points in the CNN, respectively.

V. EXPERIMENTAL RESULTS

*A. Reuse-aware shortcut optimizer*

Fig. 16(a), and 16(b) show the buffer size, DRAM access and latency (i.e., inference time per single image) with regard to the cut-point position for YOLO v2. The minimum SRAM required is 0.76 MB corresponding to layer 12 (CONV9). Compared to the *baseline* which uses a fixed row-based weight reuse scheme, the proposed scheme achieves a 2.17 times speed-up, as shown in Fig. 16(c), while requiring a 5.73 times smaller buffer size. The speed up is seen from CONV9 because the proposed scheme reuses feature-maps on-chip and the weight load time is hidden by frame-based computation (as shown in Fig. 3(a)).

Fig. 17 shows the performances of the various CNNs: YOLO v3 (77 CONV layers), ResNet152 (152 CONV layers), and EfficientNet-B1 (139 CONV layers) with respect to the switching point positions. The optimizer provides an exhaustive search to find the minimum buffer size that satisfies the DRAM access constraints. It should be noted that in Fig. 17, the weights are accessed exactly once. Hence, the total DRAM access is

$$DSP\ efficiency = \frac{Average\ GOPS}{Peak\ GOPS} = \frac{Average\ GOPS}{4 \times freq \times N_{DSP\ used}}$$

TABLE II
RESNET152 - PERFORMANCE COMPARISON TO PREVIOUS WORKS

| Features | HPCA'19 [8] | Proposed |
|---|---|---|
| FPGA board | VC707 | KCU1500 |
| Frequency | 150 MHz | 200 MHz |
| Logics (K) | 283.8 (86%) | 215.3 (33%) |
| DSPs | 2800 (100%) | 2240 (41%) |
| BRAM18K | 2040 (99%) | 1945 (45%) |
| Input size | 224x224 | 224x224 |
| CNN size (GOP) | 22.63 | 23.86 |
| Precision | 16-bit | 16-bit |
| Weights (MB) | 112.6 MB | 112.6 MB |
| Latency (ms) | 35.24 | 39.27 |
| Throughput | 608.3 GOPS | 607.5 GOPS |
| DSP efficiency | 72.4% | 71.1% |
| Weight Load | Multiple times | Once |
| Off-chip FMs | 62.93 MB | 11.97 MB |

TABLE III
MINIMUM BUFFER SIZE FOR EACH CNN

| Networks | Input size | Number of layers[*] | Minimum required buffer size |
|---|---|---|---|
| YOLO v2 | 416x416 | 21 | 0.762 MB |
| VGG-CONV | 224x224 | 13 | 0.712 MB |
| YOLO v3 | 416x416 | 106 | 1.682 MB |
| RetinaNet | 512x512 | 137 | 2.392 MB |
| Resnet50/152 | 224x224 | 68/204 | 1.039 MB |
| EfficientNet-B1 | 256x256 | 181 | 0.43 MB |

[*] including other layers such as shortcut, concatenation, etc.

TABLE IV
BUFFER SIZE TO MINIMIZED OFF-CHIP ACCESS FOR VGG-CONV

| | OLAccel[38] | SmartShuttle[12] | Proposed |
|---|---|---|---|
| Networks | | VGG-CONV | |
| Precision | Mixed (4,8) | 8-bit | 8-bit |
| SRAM size | 2.4 MB | 0.75 MB | 0.712 MB |
| DRAM access | 42.8 MB | 58.1 MB | 42.8 MB |

Where 4 is the number of INT8 operations that a single DSP can do in a single cycle, and *freq* is the clock frequency.

Table II presents the comparison of the proposed design over the previous work on the ResNet152 inference. For a fair comparison, the proposed accelerator is designed with 16-bit precision, and the BRAMs constraint is set similar to Shortcut Mining in [8]. In this case, each multiplication is mapped to a single DSP. The proposed scheme achieves a similar DSP efficiency while reducing the off-chip access for the weights and feature-maps significantly. Shortcut Mining uses a large number of parallel buffer banks which are shared for both the feature-maps and partial sums. Because the bit width of the partial sums is many fold larger than that of the feature-maps, some of the buffer space might be wasted. In addition to that, a fixed data reuse scheme in [8] might require a very high BRAM utilization and frequent off-chip access.

*B. Minimum buffer size requirement to satisfy the DRAM access constraints*

Table III shows the minimum required buffer size for various CNNs to meet the DRAM access constraints (equation 10). These buffer sizes are not only practical for small to medium size FPGA chips but also ASIC chips where the size of the SRAM might dictate the chip size. For example, the Google TPU [28] consists of 28 MiB of on-chip buffer which accounts for 30% of the chip area. To efficiently use the proposed design flow on ASIC design, three physical buffers need to be merged to a unified buffer. Accordingly, the controller and some data paths needs to be modified to support a unified buffer. Moreover, Table II shows that the minimum required buffer size for various CNNs is practical for ASIC chip. Hence, with a same design flow, the switching points of the proposed data reuses can be move around to adjust the off-chip access based on the available on-chip memory.

The DRAM access constraints limit the DRAM accesses which are also important to reduce the energy consumption for ASIC chips [37]. Table IV compares the buffer size and DRAM access of the proposed scheme compared to previous works for always larger than the DRAM access for the feature-maps (FM) by the amount of the weight size. It can be seen that for all CNNs, the cut-point at the beginning achieves a better latency at the cost of a larger buffer size. As long as the buffer constraints are satisfied, the frame-based weight reuse scheme is better than the row-based weight reuse scheme in terms of both latency and DRAM access reduction.

The DSP efficiency (i.e., MAC efficiency for designs in which MACs are inferred to DSPs) is calculated as follows:

TABLE V
PERFORMANCE OF THE VARIOUS CNNS USING THE PROPOSED SCHEME

| | ResNet50 | ResNet152 | Yolo v2 | YOLO v3 | RetinaNet | EfficientNet-B1 |
|---|---|---|---|---|---|---|
| Platform | Xilinx KCU1500 | | | | | |
| Frequency | 200 MHz | | | | | |
| Data format | 8-bit | | | | | |
| Input size | 256x256 | 256x256 | 416x416 | 416x416 | 512x512 | 256x256 |
| CNN Size (GOP) | 11.76 | 31.16 | 17.18 | 65.86 | 102.2 | 1.38 |
| LUTs/FFs (K) | 212.7/361.5 | 212.7/361.5 | 203.1/331.0 | 213.3/352.0 | 264.3/367.2 | 264.1/375.7 |
| DSP utilization | 2240 | 2240 | 2240 | 2240 | 2240 | 2240 |
| BRAM18k | 2368 (55%) | 2368 (55%) | 2304 (53%) | 3020 (70%) | 3766 (87%) | 2594 (50%) |
| Latency (ms) | 11.69 | 26.78 | 14.73 | 57.57 | 93.16 | 4.69 |
| Frame rate (fps) | 85.5 | 37.3 | 67.9 | 17.4 | 10.7 | 213.2 |
| GOPS | 1006 | 1163 | 1166 | 1142 | 1097 | 317.1 |
| MAC Efficiency | 61.4% | 71.0% | 71.2% | 69.7% | 67.0% | 19.37% |
| Weight load | Once | Once | Once | Once | Once | Once |
| Off-chip FMs | **0.19 MB** | **0.19 MB** | **0.66 MB** | **90.6 MB** | **136.4 MB** | **0.19 MB** |
| Total off-chip [*] | 59.09 MB | 130.2 MB | 48.9 MB | 153.5 MB | 261.34 MB | 60.7 MB |
| Off-chip reduction | **60.62%** | **56.7%** | **70.31%** | **60.34%** | **47.81%** | **84.81%** |

[*] Total off-chip memory access if data (weights/inputs/outputs) are accessed exactly once.

TABLE VI
COMPARISON OF END-TO-END FRAMEWORKS ON RESNET50 INFERENCE

|  | ML-Suite [44] | FPL'19 [33] | Cloud-DNN [17] | Proposed |
|---|---|---|---|---|
| Platform | VU9P (16nm) | VU9P (16nm) | VU9P (16nm) | KCU1500 (20nm) |
| URAM(*) size | 270 Mb | | | -- |
| Frequency | 500 MHz | 125 MHz | 214 MHz | 200 MHz |
| Framework | Tensorflow | Tensorflow | Caffe | Tensorflow |
| Network | ResNet50 | | | |
| Input size | 224x224 | 224x224 | 224x224 | 256x256 |
| Precision | 8-bit | 8-bit | 16-bit | 8-bit |
| Latency | 7.77ms | 23.8ms | 8.12ms | 11.9ms |
| LUTs | 612K | 605K | 696K | 217K |
| DSPs | 5493 | 6005 | 5489 | 2240 |
| GOPS | 1290 | 328 | 1235 | 1006 |
| Data reuse | Fixed | Fixed | Fixed | **Flexible** |
| Shortcut reuse & fusion in HW | No | No | No | **Yes** |
| SRAM size (MB) | 31.2 [*] | 18.8 [*] | 38.3 [*] | **5.2** |
| DSP efficiency | 23.47% | 21.85% | 52.58% | **56.14%** |

[*]: URAM (Ultra-RAM) + BRAMs utilization

TABLE VII
EFFICIENTNET-B1 INFERENCE PERFORMANCE ON THE PROPOSED DESIGN

| Resolution | 256×256 | 512×512 | 768×768 |
|---|---|---|---|
| FPGA board | KCU1500 | | |
| Frequency | 200 MHz | | |
| LUTs/FFs (K) | 264.1/375.7 | 264.5/375.5 | 271.7/375.4 |
| DSPs | 2176 | | |
| BRAM18Ks | 2594 (60%) | 2723 (62%) | 3845 (89%) |
| GOPS | 317.1 | 267.4 | 274.4 |
| DSP efficiency | 19.37% | 16.3% | 16.75% |
| Off-chip FMs | 0.19 MB | 144 MB | 344 MB |
| Total off-chip [*] | 60.7 MB | 216 MB | 475 MB |
| Off-chip Weights | Once | | |
| Off-chip reduction | **84.81%** | **29.2%** | **27.6%** |
| Power (W) | 21.09 | 23.76 | 26.71 |
| GOPS/W | 15.0 | 11.3 | 10.3 |

[*]: Total off-chip memory access if weights/inputs/outputs are accessed from DRAM exactly once.

chip access for intermediate data including the shortcut data while using less SRAM resource. Compared to Cloud-DNN, the proposed work utilizes 7.4× less SRAM resource while having 1.07× higher DSP efficiency. In addition, the proposed work achieves a competitive GOPS and 2.4× higher DSP efficiency than ML-Suite while requiring 6.0× less SRAM resource and running at 2.5× lower frequency.

*C. Scalability and power efficiency for SOTA CNNs*

A larger input size leads to higher accuracy while causing an increase in the on-chip buffer requirement, DRAM access and latency. Table VII presents the performance of the EfficientNet-B1 inference with various high resolution images to demonstrate the scalability of the proposed scheme. For example, with a 768×768 input size, the total DRAM access if the inputs/outputs are accessed exactly once is 475 MB. Out of this, the proposed scheme requires only 344 MB for the feature-maps access which results in a 27.6% reduction. The reduction is 29.2% and 84.81% for 512×512 and 256×256, respectively.

the VGG16 CONV layers. With the same amount of DRAM access (i.e., input/output are accessed once), the proposed scheme requires a 3.4 times smaller buffer than [38] due to the adaptive reuse policy. Compared to SmartShuttle which proposes layer-wise data reuse schemes, the proposed scheme reduces DRAM access by 1.36 times with smaller on-chip buffer size. This demonstrates the efficiency of adaptive switching between row-based weight reuse schemes and frame-based weight reuse schemes over the reuse schemes using the tiled input/output in SmartShuttle.

Finally, the performance of various state-of-the-art CNNs using the proposed scheme are shown in Table V. Depending on the on-chip buffer constraints, the proposed scheme minimizes the DRAM access for the feature-maps while accessing parameters exactly once for all the CNNs due to strict off-chip access constraints in the proposed optimization. Compared to the baseline, in which all data are accessed off-chip exactly once, the proposed scheme reduces the total DRAM access by 47.8-84.8% for the various CNNs.

Table VI shows the comparison of some end-to-end frameworks for the ResNet50 inference. All three previous works utilize large Ultra-RAM of a cloud-scale Xilinx FPGA which has 6840 DSPs and 270 Mb of Ultra-RAM. However, none of them support flexible data reuse. Whereas, the proposed framework supports adaptive data reuse schemes with in-hardware shortcut fusion, thereby completely removing the off-

The power of the accelerator is estimated as the sum of FPGA-chip power plus the DRAM power. The FPGA-chip power is calculated by Xilinx Power Estimator with the signal switching frequency from RTL simulation. The DRAM access energy is estimated from the total DRAM access and the energy per access from [56]. The CPU for experiment is Intel Xeon E3-1245 v5 3.5GHz with OpenMP enable. Meanwhile, GPU is tested with Pytorch 1.8.0 or Keras on Tensorflow 2.3.0 and CUDA 10.2. GPU power is calculated by *nvidia-smi*. Compared to the GPU performance on Keras in Fig. 2, the proposed design is 3.0×, 4.6×, and 8.6× faster than NVIDIA RTX 2080 Ti for

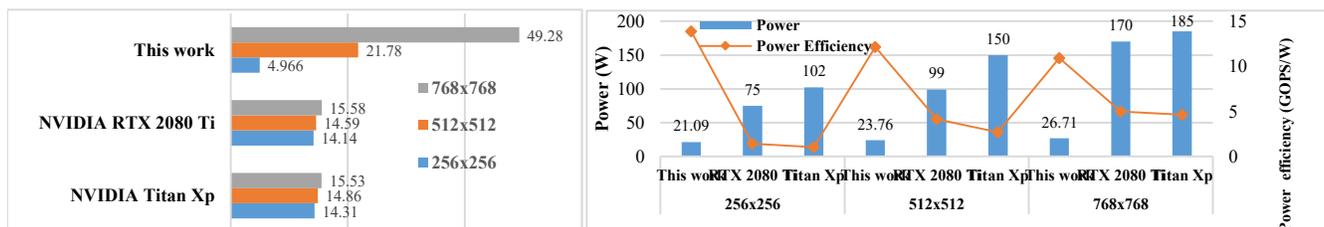

(a) Latency (ms) w/ different input size (Pytorch 1.8.0, CUDA 10.2)  (b) Power and Power efficiency
Fig. 18. Comparison of the proposed work and modern GPUs for EfficientNet-B1 inference (batch size = 1).

256×256, 512×512, and 768×768 input sizes, respectively. It is noticeable that the GPU performance on Pytorch is much higher that on Keras due to better kernel optimization. Fig. 18 shows a detailed comparison of the proposed work with GPUs on Pytorch. The proposed work is 2.8× faster than RTX 2080 Ti for 256×256. For larger input sizes, GPUs outperforms the proposed work because of better utilization of huge number of CUDA cores compared to limited parallelism of the proposed design. However, in terms of power efficiency, the proposed design is 9.9×, 2.9×, and 2.2× better than RTX 2080 for 256×256, 512×512, and 768×768 input sizes, respectively. The DSP efficiency is low for EfficientNet (e.g., less than 20%) due to low density of multiplications in depthwise convolution. However, compared to GPUs, which has huge number of parallelism, the proposed design still shows significant speed up for 256×256 input size thanks to the reuse-aware static memory allocation and shared MAC design.

## VI. RELATED WORKS

Hardware accelerators use various dataflows or optimizations to increase resource utilization, for example, weight stationary [7], [8], [16], [28], [48], [53], output stationary [49], and row stationary [13], [50], [51]. MAESTRO [52] analyzes the energy-performance trade-off for the various dataflows above to choose an optimized one for a given CNN. These fixed dataflow designs result in a sub-optimal on-chip buffer size and off-chip memory access when running different layers of a CNN with different characteristics. FlexFlow [11] presents an optimization from the on-chip buffer to the PEs enabling the mixing of multiple parallelism types of the feature-maps, neurons, and synapses to boost resource utilization which is orthogonal to the proposed work. DNA [10], and SmartShuttle [12] propose using a layer-wise data reuse scheme which supports switching between two of the three schemes: Input-Reuse, Output-Reuse and Weight-Reuse. These works reduce the off-chip access efficiently compared to previous works which have a similar global buffer size. However, in these works, a larger buffer size (i.e., 512 KB) has less benefits even though the accelerator supports a flexible tile size. The ASIC design in [57] uses a configurable dataflow for 1x1 and 3x3 convolutional layers to increase resource utilization. The study in [58] proposes an adaptive row-based weight reuse scheme opted for depthwise convolution whereas the proposed work targets for both normal, depthwise convolution, and residual layers. Finally, the research in [59] presents a hardware/software codesign to mixed non-pruned and pruned layers in CNN computation based on each layer's characteristic. Different from this work, the proposed study runs the original CNN while using a block-wise data reuse scheme to reduce off-chip memory access.

In the literature, there are some compilers to schedule a FPGA-based CNN accelerator such as TVM [43], Xilinx ML-Suite [44], Intel DLA [45], and DNNVM [46]. TVM optimizes the standard CNN inference on the software side for a vanilla deep learning accelerator by overlapping the tensor computation with the memory load/store operations. However, the machine learning-based optimization for tuning each convolution layer has a considerable time cost which is burdensome for very deep networks. DLA optimizes CNN graph by adding a 1x1 identity layer and merging element-wise addition to the previous layer. Finally, ML-Suite and DNNVM fuse many adjacent layers such as convolution, batch norm, Relu, and pooling to reduce the off-chip access for intermediate data. Nevertheless, the lack of an in-hardware flexible data reuse and shortcut reuse reduces the MAC efficiency. Whereas, the proposed hardware/software co-optimizer, while being fast, provides adaptive data reuses to minimize the off-chip memory access and improve the MAC efficiency even with the on-chip buffer constraints.

## VII. CONCLUSION

This paper presents a tool for FPGA-based CNN inference which uses a reuse-aware shortcut optimizer to minimize the latency, the off-chip memory access and improve the MAC efficiency given the on-chip buffer constraints. Comprehensive comparisons to previous works demonstrate the efficiency of the proposed approach. In addition, the proposed work achieves superior performance compared to NVIDIA GPUs when running state-of-the-art Squeeze-and-Excitation-based CNNs such as EfficientNets/EfficientDet/MobileNet v3.

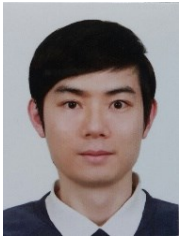

**Duy Thanh Nguyen** received the B.S. degree in electrical engineering from Hanoi University of Science and Technology, Hanoi, Vietnam, M.S., and Ph.D. degree in Electrical and Computer Engineering from Seoul National University, Seoul, Korea, in 2011, 2014, and 2021, respectively. He is currently working at Samsung Electronics, Korea.

His research interests include computer architecture, memory system, VLSI design for computer vision, hardware/software co-design for deep learning acceleration.

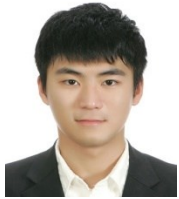

**Hyeonseung Je** received the B.S. degree in electronic engineering from Sogang University, Seoul, South Korea in 2019, and the M.S. degree in electrical and computer engineering from Seoul National University, Seoul, South Korea in 2021. He is currently a Software Engineer with Kakao in South Korea.

His current research interests include deep learning applications, recommendation systems and computer architecture.

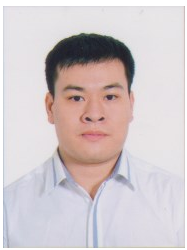

**Tuan Nghia Nguyen** received the B.S. degree in Electronics and Telecommunications from Hanoi University of Science and Technology, Hanoi, Vietnam, and the M.S. degree in Electrical and Computer Engineering from Seoul National University, Seoul, Korea in 2019. He is currently working toward the Ph.D. degree in Electrical and Computer Engineering at Seoul National University.

His research interests include computer vision, deep learning applications, and computer architecture.

**Soojung Ryu** is CEO of SAPEON KOREA Inc. and VP of SK Telecom. She is leading development of SAPEON, which is the first Korean commercialized AI accelerator chip for server. Before joining SK telecom, she was a University-Industry Collaboration Professor at Seoul National University. She worked for Samsung Electronics from 2004 to 2018 as a processor architect after she received her Ph.D. degree in Electrical & Computer Engineering from Georgia Institute of Technology. She was a Samsung Corporate Vice President from Dec 2014 to 2018. She was leading a processor design and SW framework development for digital signal processor (Samsung Reconfigurable Processor) and Samsung Mobile Graphics Processing Unit. Her current research focus is deep neural network accelerator design as well as the high-performance processor solutions for various application areas.

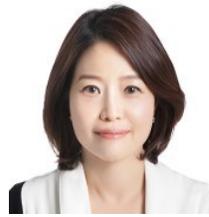

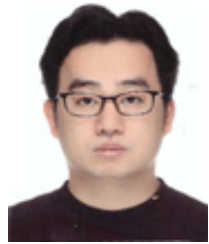

**Kyujoong Lee** received the B.S. degree in electrical engineering from Seoul National University, Seoul, Korea, in 2002 and the M.S. degree in electrical engineering from University of Southern California, Los Angeles, USA, in 2008. He got Ph.D. degree in electrical engineering and computer science at Seoul National University Seoul, Korea, in 2013. From 2002 to 2005, he was with Com2us corporation, Seoul, Korea, as a developer. From 2013 to 2017, he worked for S.LSI division of Samsung Electronics corporation. In 2017, he was appointed as an Assistant Professor in the department of Electronic Engineering at Sun Moon University, Asan, Korea. In 2022, he moved as an Associated Professor to the school of AI Convergence at Sungshin Women's University, Seoul, Korea. His major research interests include the algorithms and architectures of deep learning and image/video processing.

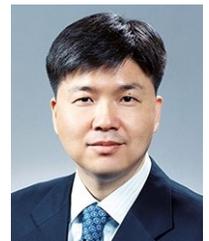

**Hyuk-Jae Lee** received the B.S. and M.S. degrees in electronics engineering from Seoul National University, South Korea, in 1987 and 1989, respectively, and the Ph.D. degree in Electrical and Computer Engineering from Purdue University, West Lafayette, IN, in 1996. From 1996 to 1998, he was with the Faculty of the Department of Computer Science, Louisiana Tech University, Ruston, LS. From 1998 to 2001, he was with the Server and Workstation Chipset Division, Intel Corporation, Hillsboro, OR, as a Senior Component Design Engineer. In 2001, he joined the School of Electrical Engineering and Computer Science, Seoul National University, South Korea, where he is currently a Professor. He is a Founder of Mamurian Design, Inc., a fabless SoC design house for multimedia applications. His research interests are in the areas of computer architecture and SoC design for multimedia applications.